\documentclass[]{spie}  

\usepackage{amsmath,amsfonts,amssymb}
\usepackage{graphicx}
\usepackage[colorlinks=true, allcolors=blue]{hyperref}

\title{High Contrast and High Angular Imaging at Subaru Telescope}

\author[a,b,c,d]{Olivier Guyon}
\author[a]{Kyohoon Ahn}
\author[e]{Masayuki Akiyama}
\author[a]{Thayne Currie}
\author[a]{Vincent Deo}
\author[a]{Takashi Hattori}
\author[a,d]{Tomoyuki Kudo}
\author[a]{Julen Lozi}
\author[a]{Yosuke Minowa}
\author[a]{Yoshito Ono}
\author[a,f]{Nour Skaf}
\author[d,g,h]{Motohide Tamura}
\author[a]{Vincent Vievard}

\affil[a]{Subaru Telescope, National Astronomical Observatory of Japan, National Institutes of Natural Sciences (NINS), 650 North A`oh\={o}k\={u} Place, Hilo, HI 96720, United States}
\affil[b]{Steward Observatory, University of Arizona, Tucson, AZ 87521, United States}
\affil[c]{College of Optical Sciences, University of Arizona, Tucson, AZ 87521, United States}
\affil[d]{Astrobiology Center, 2 Chome-21-1, Osawa, Mitaka, Tokyo, 181-8588, Japan}
\affil[e]{Astronomical Institute, Tohoku University, Aramaki, Aoba, Sendai 980-8578, Japan}
\affil[f]{LESIA, Observatoire de Paris, Univ.~PSL, CNRS, Sorbonne Univ., Univ.~de Paris, 5 pl. Jules Janssen, 92195 Meudon, France}
\affil[g]{National Astronomical Observatory of Japan, 2-21-2, Osawa, Mitaka, Tokyo 181-8588, Japan}
\affil[h]{Department of Astronomy, Graduate School of Science, The University of Tokyo, 7-3-1, Hongo, Bunkyo-ku, Tokyo, 113-0033, Japan}

\authorinfo{Further author information: (Send correspondence to O.G.)\\O.G.: E-mail: guyon@naoj.org}

\begin{document} 
\maketitle

\begin{abstract}
Adaptive Optics projects at Subaru Telescope span a wide field of capabilities ranging from ground-layer adaptive optics (GLAO) providing partial correction over a 20 arcmin FOV to extreme adaptive optics (ExAO) for exoplanet imaging. We describe in this paper current and upcoming narrow field-of-view capabilities provided by the Subaru Extreme Adaptive Optics Adaptive Optics (SCExAO) system and its instrument modules, as well as the upcoming 3000-actuator upgrade of the Nasmyth AO system. 
\end{abstract}

\keywords{Adaptive Optics, High Contrast Imaging, Coronagraphy, Exoplanets}

\section{Adaptive Optics at Subaru Telescope: Overview}
\label{sec:AOoverview}  

Adaptive optics at the Subaru Telescope \cite{2020SPIE11448E..0KO} started with the 36-element curvature system at the telescope's Cassegrain focus \cite{2004PASJ...56..225T}. A more capable 188-element system with both LGS and NGS modes was deployed in the 2000s at the telescope's infrared Nasmyth focus \cite{2008SPIE.7015E..10H,2010SPIE.7736E..0NH}, and is still in operation, feeding the Infrared Camera and Spectrograph (IRCS) \cite{1998SPIE.3354..512T, 2004SPIE.5492.1542T, 2000SPIE.4008.1056K} and Subaru Coronagraphic Extreme AO (SCExAO)\cite{2015PASP..127..890J,2020SPIE11448E..0NL} instruments.

New adaptive optics capabilities are currently in development, focusing on three major directions :
\begin{itemize}
\item{{\bf Ground-Layer Adaptive Optics (GLAO)}, providing $\approx$ 0.2-arcsec image quality over a 20-arcmin field of view at the telescope's Cassegrain focus using laser guide stars (LGSs) and an adaptive secondary mirror. GLAO development is the core part of the ULTIMATE\cite{2020SPIE11203E..0GM,2020SPIE11450E..0OM} project currently in development phase, with first light anticipated in 2028.}
\item{{\bf Laser Tomography Adaptive Optics (LTAO)}, delivering diffraction-limited imaging over $\approx$ 20-arcsec field of view at the telescope's IR Nasmyth focus using LGSs for full sky coverage. LTAO is part of the ULTIMATE project, and will be deployed in year 2023 at the IR Nasmyth platform with the ULTIMATE-START\cite{2020SPIE11448E..1OA} project.}
\item{{\bf Extreme Adaptive Optics (ExAO)} providing high image quality for high contrast imaging and visible-light diffraction limited imaging over small field of views. ExAO is implemented in the Subaru Coronagraphic Extreme AO (SCExAO) platform and its instrument modules.}
\end{itemize}

We describe in this paper the current and near-future capabilities of narrow-field adaptive optics modes, with a focus on ExAO and including LTAO, and their implementation on the telescope's Nasmyth IR (NasIR) platform. We describe in \S \ref{sec:NasIR} the NasIR instrument configuration, including the upcoming deployment of the NasIR beam switcher, which allows for multiple instrument to be AO-fed. Observing capabilities are discussed in \S \ref{sec:obsmodes}. The main development activities and future perspectives are discussed in \S \ref{sec:dev}.

\section{Nasmyth IR Platform Optical Configuration}
\label{sec:NasIR}

\subsection{Current Configuration}

\begin{figure}[ht]
    \centering
    \includegraphics[width=17cm]{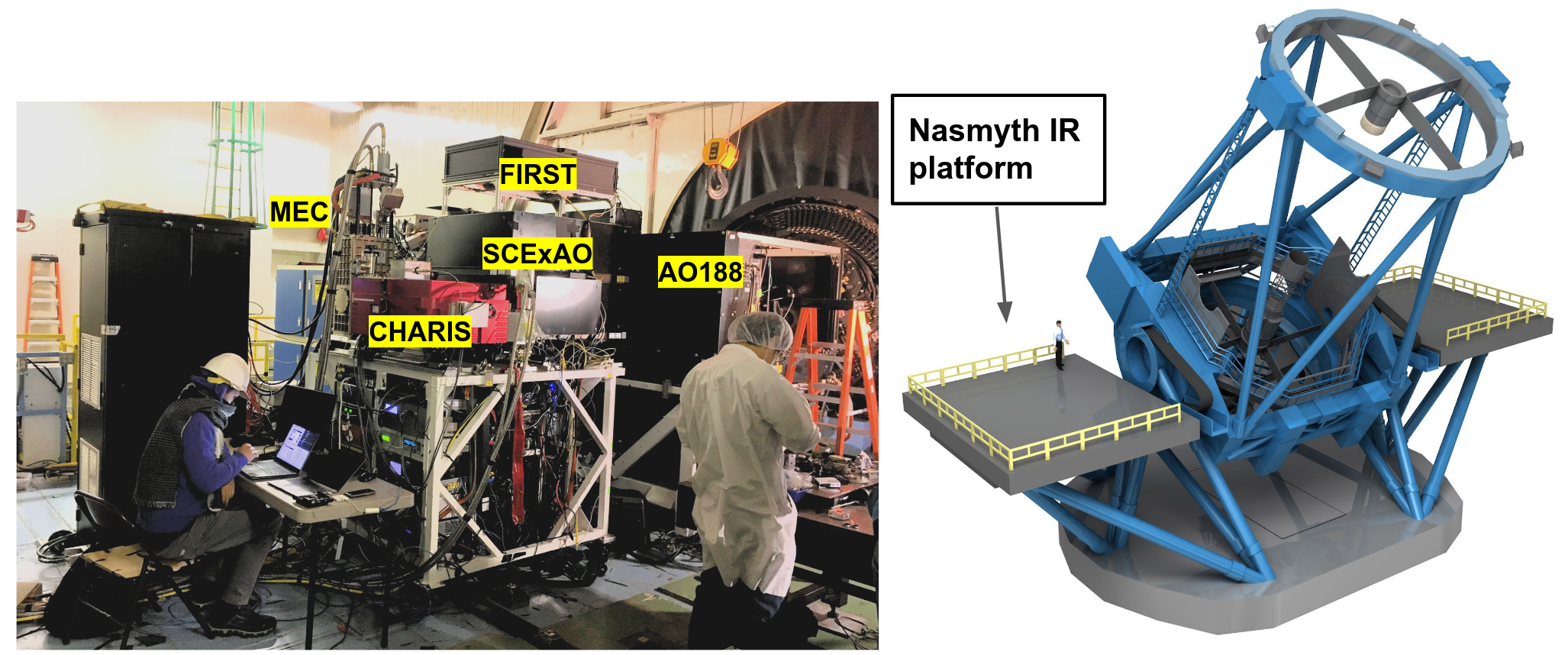}
    \vspace*{0.3cm}
    \caption{Subaru Telescope's infrared Nasmyth (NasIR) platform hosts its narrow-field adaptive optics instrumentation. The NasIR platform is one of two large gravity-invariant platforms on the telescope's side (right). A recent picture of the platform (left) shows the AO188 and SCExAO instrument configuration. Both images are in approximately the same 3D orientation.}
    \label{fig:NasIR}
\end{figure}

Narrow-field AO instrumentation is hosted at the telescope's NasIR platform, as shown in Figure \ref{fig:NasIR}. The current 188-element facility AO system provides NGS and LGS correction for the IRCS and SCExAO instruments. IRCS provides diffraction-limited imaging and spectroscopy in near-IR while SCExAO includes a second-stage Extreme AO correction with a 2000-element MEMS deformable mirror.

The IRCS and SCExAO observing modes are mutually exclusive and cannot be scheduled within the same night. Switching between both instruments requires physically swapping the instruments between the storage and in-focus locations. The picture in Figure \ref{fig:NasIR} shows the SCExAO configuration, while the IRCS configuration would have IRCS at the AO188 focus. Switching between the two modes is done in daytime with a overhead crane (partially visible in Figure \ref{fig:NasIR} above AO188). This limitation will be addressed by deploying the Nasmyth beam switcher (NBS) behind the facility AO system.

\subsection{Beam Switcher Configurations}

The Nasmyth bean switcher \cite{Zheng2022} (NBS) will be deployed by 2024 to allow for on-the-fly switching between SCExAO and IRCS without requiring instrument craning, and will also support expanding the suite of AO-fed instruments. The NBS also allows for simultaneous operation of instrument modules by using dichroic beam splitters.

\begin{figure}[ht]
    \centering
    \vspace*{-0.1cm}
    \includegraphics[width=15cm]{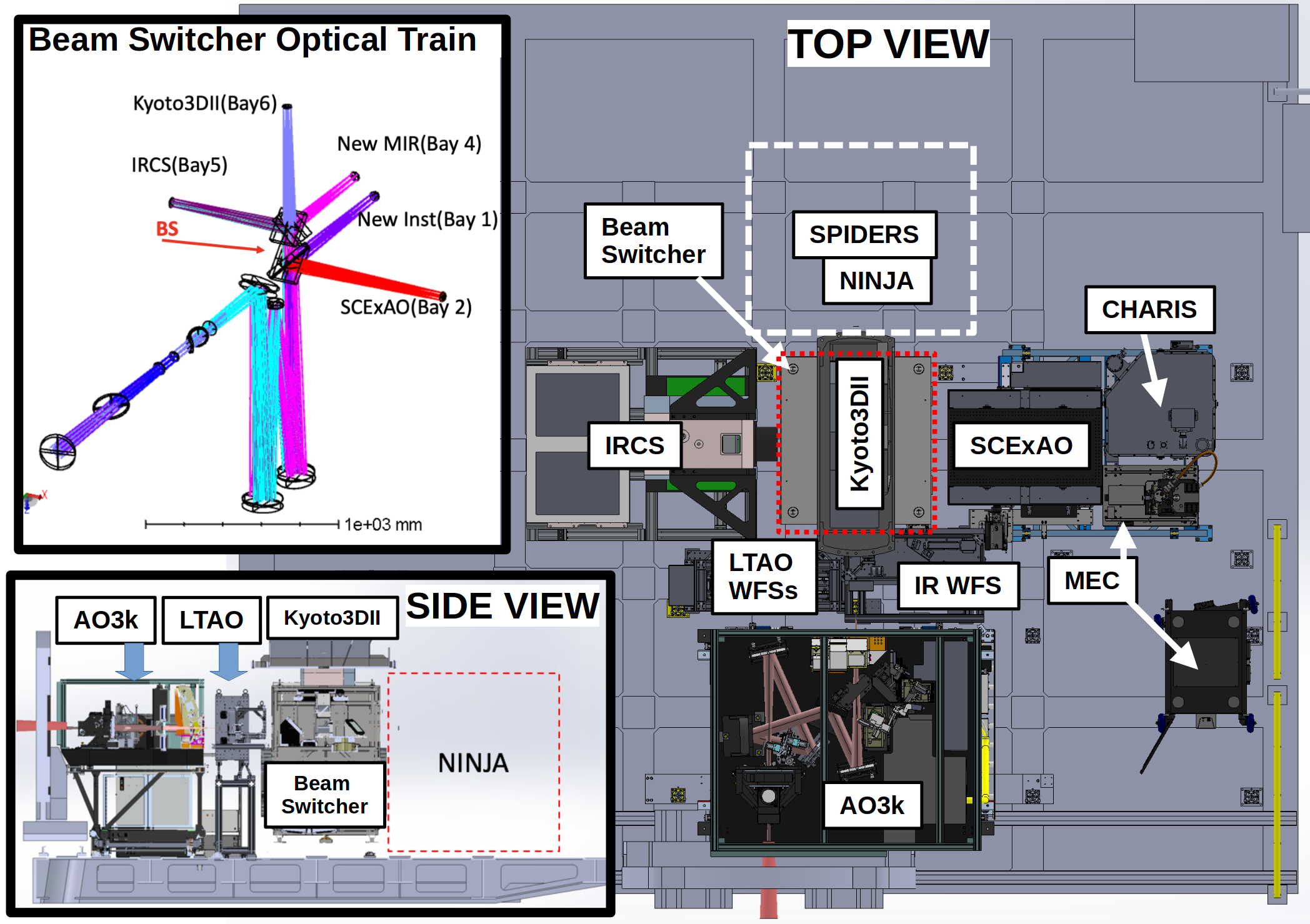}
    \includegraphics[width=10cm]{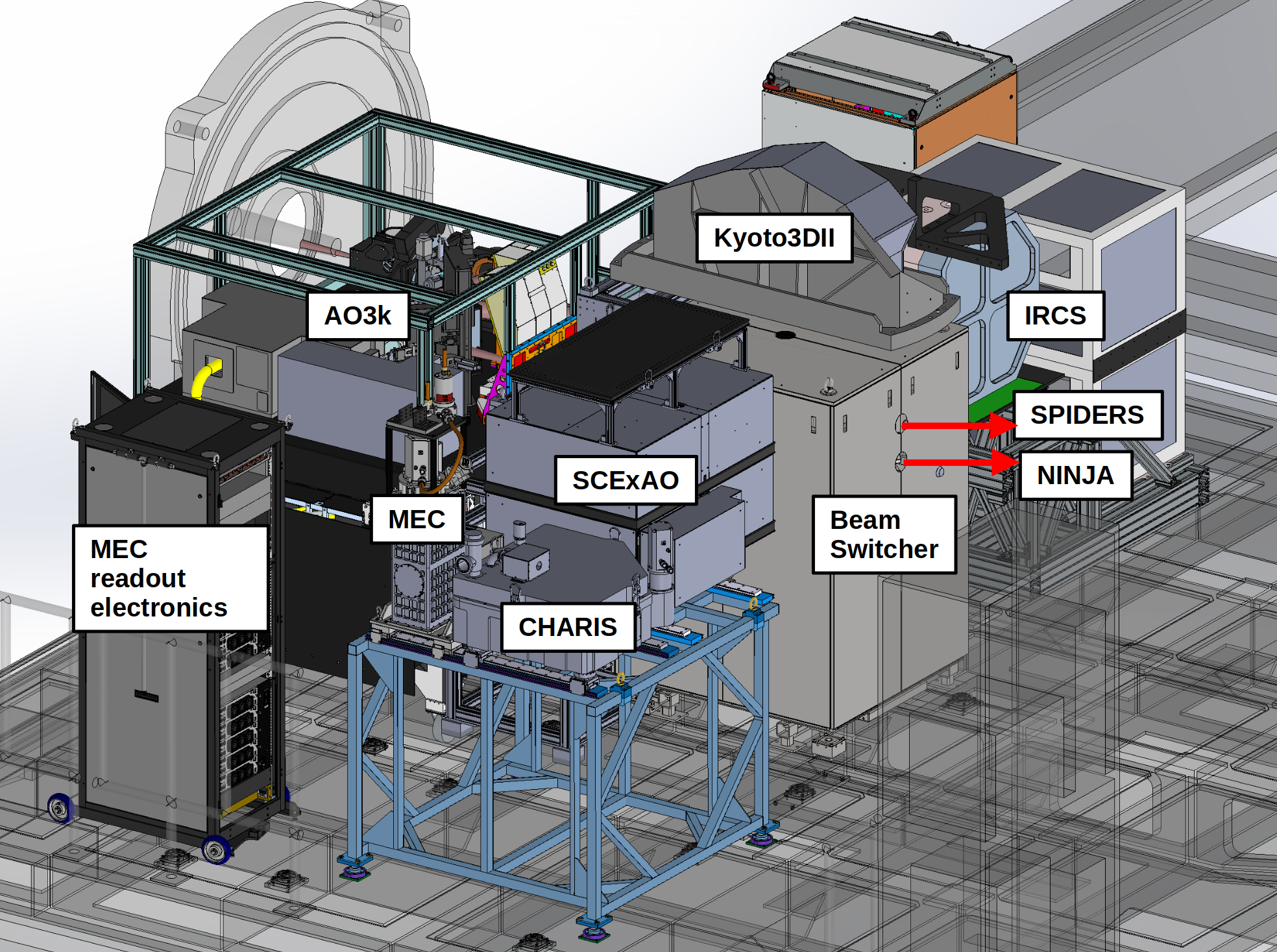}
    \caption{NasIR platform layout with beam switcher. In the top view, light enters the platform from the bottom into AO3k. The Nasmyth beam switcher (NBS) can feed instruments on 4 output ports (left: IRCS, right: SCExAO, through: SPIDERS or NINJA, top: Kyoto3DII). A perspective view is shown in the bottom.}  
    \label{fig:NasIR-top}
\end{figure}

\begin{figure}[ht]
    \centering
    \includegraphics[width=17cm]{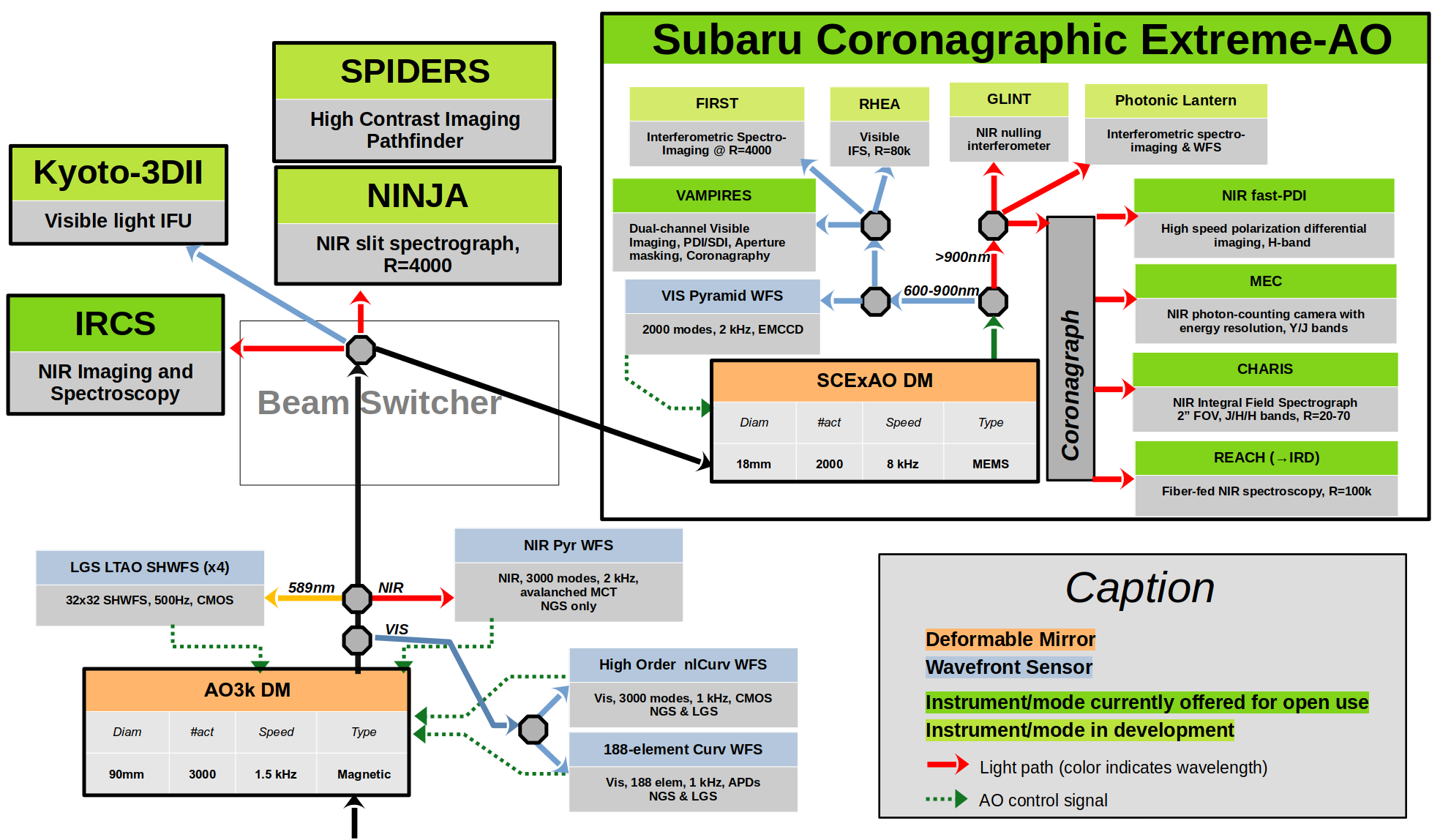}
    \vspace*{0.3cm}
    \caption{Adaptive Optics instruments and overall beam path. Instrument modules currently available for open use include IRCS, SCExAO/VAMPIRES, SCExAO/CHARIS, SCExAO/MEC, SCExAO/FastPDI, and REACH/IRD. The NINJA, SCExAO/GLINT, SCExAO/FIRST and AO3k/Kyoto3DII are in development phase and be available to observers in the near future. SPIDERS, SCExAO/RHEA and SCExAO/PhotonicLantern are currently focused on technology validation of novel instrument concepts.}
    \label{fig:AOblockdiag}
\end{figure}

Figure \ref{fig:NasIR-top} shows the NasIR instrument platform layout, and Figure \ref{fig:AOblockdiag} shows the light path through adaptive optics key elements and instruments.

\section{Observing Modes and Capabilities}
\label{sec:obsmodes}

\subsection{Adaptive Optics Correction}

All adaptive optics instrumentation at NasIR is first corrected by the AO3k system \cite{Lozi2022}. AO3k is an upgrade of the current 188-element system, where the 188-actuator bimorph DM is replaced by a 64x64 magnetic DM (model ALPAO DM3228), and high order visible and NIR wavefront sensors are added. AO3k supports the following modes:
\begin{itemize}
    \item Visible light NGS single conjugated AO (NGS-VIS-SCAO)
    \item Near-infrared light NGS single conjugated AO (NGS-NIR-SCAO)
    \item LGS Single conjugated AO (LGS-SCAO)
    \item Laser tomographic AO (LTAO)
\end{itemize}

The high order visible light high-order WFS will be a non-linear curvature WFS \cite{2010PASP..122...49G}, located within the AO3k enclosure. The NIR pyramid WFS and LTAO LGS Shack-Hartman WFSs are located at the output of AO3k, before the NBS.

Second-stage correction is performed by SCExAO with a 2000-actuator MEMS type mirror (model: Boston Micromachines 2k, to be upgraded in 2024 to a segmented 3k device), which provides high-speed high precision wavefront control for high contrast imaging and visible-light AO.

\subsection{Near-IR instrumentation}

The AO3k NIR corrected beam can be fed to the IRCS camera and spectrograph or the SCExAO NIR modules.

IRCS provides wide field imaging (20mas or 50mas FOV) and spectroscopy over the full NIR wavelength range from J-band to M-band. The instrument supports both low-resolution GRISM spectroscopy and high resolution (R$\approx$ 20k) echelle spectroscopy. IRCS's relatively wide field of view is a good match to the upcoming LTAO correction mode.

\begin{figure}[ht]
    \centering
    \includegraphics[width=17cm]{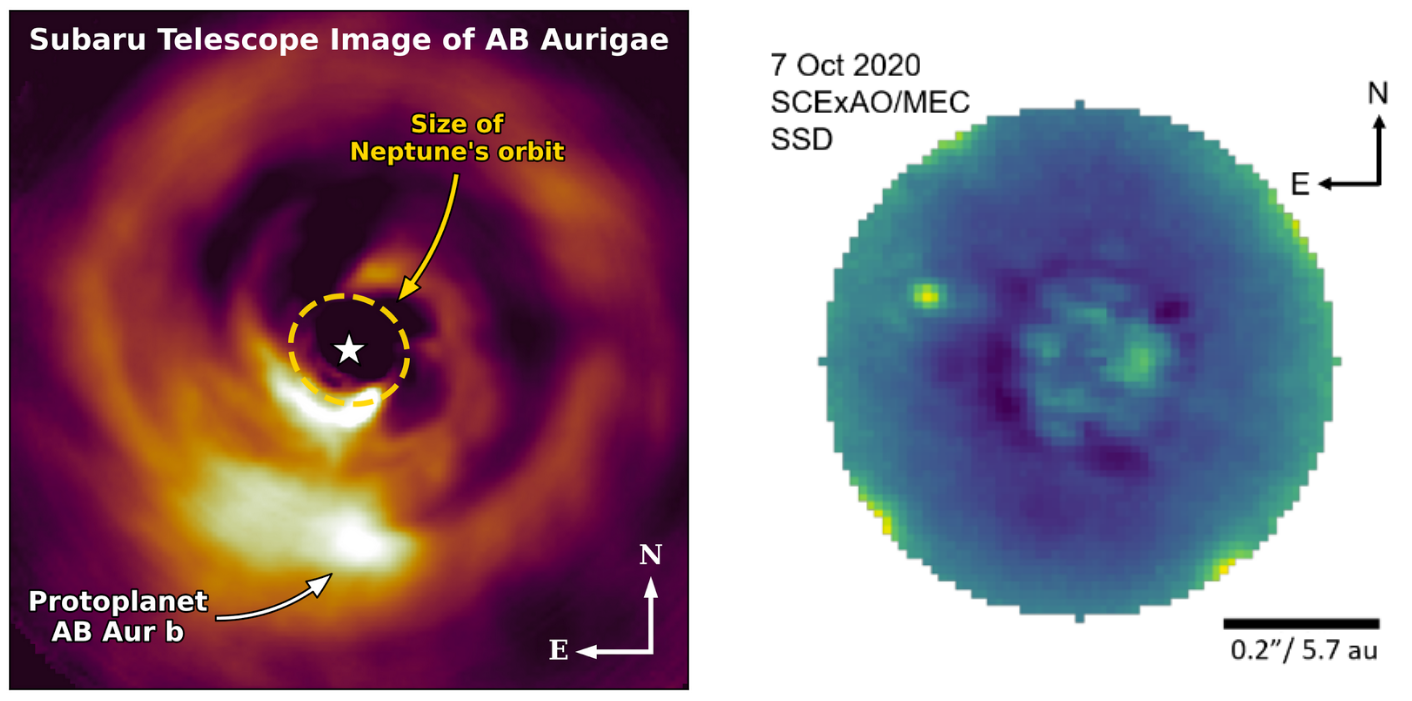}
    \vspace*{0.3cm}
    \caption{Example NIR high contrast imaging observations with SCExAO. Left: Discovery image of protoplanet AB Aur b\cite{Currie2022} with SCExAO/CHARIS. Right: Discovery image of HIP 109427 b \cite{Steiger2021} with SCExAO/MEC using stochastic speckle discrimination.}
    \label{fig:NIRimaging}
\end{figure}

Narrower field of view capabilities are provided by SCExAO instrument modules which benefit from ExAO correction. Four high contrast imaging instrument modules are currently available for open use, all compatible with SCExAO's coronagraphy mode :
\begin{itemize}
    \item The {\bf CHARIS}\cite{2015SPIE.9605E..1CG} integral field spectrograph, providing R=20 to 70 spectro-imaging over a 2 arcsecond field of view.
    \item The {\bf Fast-PDI}\cite{2020SPIE11448E..7CL} H-band high-speed polarimetric imaging camera
    \item The {\bf MKIDS exoplanet camera (MEC)}\cite{2020PASP..132l5005W}, a high speed photon-counting energy resolving camera
    \item The {\bf REACH high resolution spectroscopy}\cite{2020SPIE11448E..78K} mode proving R $=$ 100k NIR spectroscopy of exoplanets and companions
\end{itemize}
Together, these four modes provide exoplanet and disk detection and characterization by imaging, spectroscpy and polarimetry in NIR \cite{Currie2022,Kuzuhara2022,Steiger2021,Lawson2020,Lawson2021,Currie2020,Chaushev2022}. Example science observations are shown in Figure \ref{fig:NIRimaging}.

The GLINT \cite{2020MNRAS.491.4180N, 2021NatCo..12.2465M} instrument, currently in development, is a NIR nulling interferometer providing access to exoplanet and faint sources in the 0.5 to 2 $\lambda / D$ separation range. GLINT will extend high contrast imaging capabilities to smaller angular separations than possible with conventional coronagraphy.

A photonic lantern is under development for exoplanet imaging\cite{Lin2022} and high precision spectro-astrometry\cite{Kim2022}.

\subsection{Visible Light instrumentation}

Visible-light instrumentation includes the SCExAO/VAMPIRES dual-band imager, the SCExAO/FIRST spectro-interferometer and the Kyoto3DII integral field spectrograph. Kyoto3DII\cite{2010PASP..122..103S} was previously used behind AO188 and will be re-deployed on the top port of the NBS. It will benefit from AO3k correction in NGS, LGS and LTAO modes, which will provide near-diffraction-limited imaging performance in visible light with full sky coverage.

\begin{figure}[ht]
    \centering
    \includegraphics[width=17.5cm]{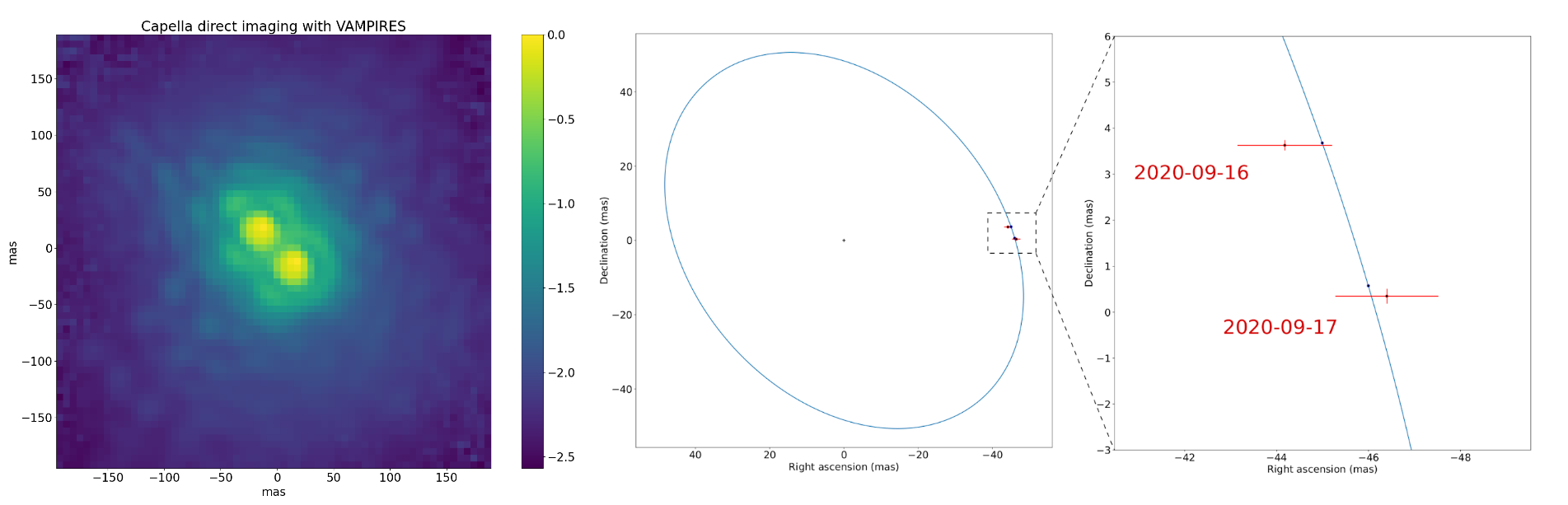}
    \vspace*{0.3cm}
    \caption{Visible-light imaging of the Capella double star with SCExAO/VAMPIRES (left) and SCExAO/FIRST (right). The two components, separated by $\approx$ 40mas, are easily resolved with VAMPIRES's imaging mode. With FIRST's sub-mas astrometric accuracy, orbital motion can be detected within hr-timescale, as shown with measurements acquired on two consecutive nights (right).}
    \label{fig:Capella}
\end{figure}

SCExAO's extreme-AO correction delivers high quality diffraction-limited imaging at visible wavelengths for the VAMPIRES and FIRST instruments. VAMPIRES\cite{2015MNRAS.447.2894N} is available for open use observations, and supports polarimetric differential imaging \cite{2020SPIE11203E..0SN, Safonov2022}, H-$\alpha$ differential imaging \cite{2020JATIS...6d5004U} and coronagraphic imaging \cite{Lucas2022}. VAMPIRES routinely operates at the diffraction limit, providing sub-20mas angular resolution, as illustrated in Figure \ref{fig:Capella} (left). 

The FIRST\cite{2020SPIE11446E..29V} fiber-fed spectro-interferometer, under development, provides high precision imaging below the telescope's diffraction limit, as illustrated on Figure \ref{fig:Capella} (right). FIRST is being upgraded with a stable compact integrated optics device feeding a R=4000 spectrograph designed for H-$\alpha$ spetro-imaging \cite{2021sf2a.conf..135L, Barjot2022, Martin2022, Martin2022b, Lallement2022}.

\section{Conclusions and Perspectives}
\label{sec:dev}

The AO188 system at Subaru Telescope feeds light to a wide range of visible and NIR instruments for imaging, spectroscopy and polarimetry. The upcoming upgrade of the 188-element system into a 3000-element system supporting NIR wavefront sensing, LGS and LTAO modes will provide significantly improved image quality over a wide range of targets. The Nasmyth beam switcher will allow efficient and simultaneous operation of multiple instruments modules, and support additional observing  modes.

Adaptive Optics developments at Subaru Telescope are both expanding scientific capabilities and prototyping new instrument concepts and techniques for future efforts. The ULTIMATE-Start LTAO effort is validating wide-field AO sensing and control solutions for the more ambitious ULTIMATE Subaru project. In high contrast imaging, SCExAO is prototyping techniques for future exoplanet-imaging systems to be installed on 30-m class telescopes \cite{2020SPIE11448E..0NL}, including the Thirty Meter Telescope (TMT).  

\begin{figure}[ht]
    \centering
    \includegraphics[width=17cm]{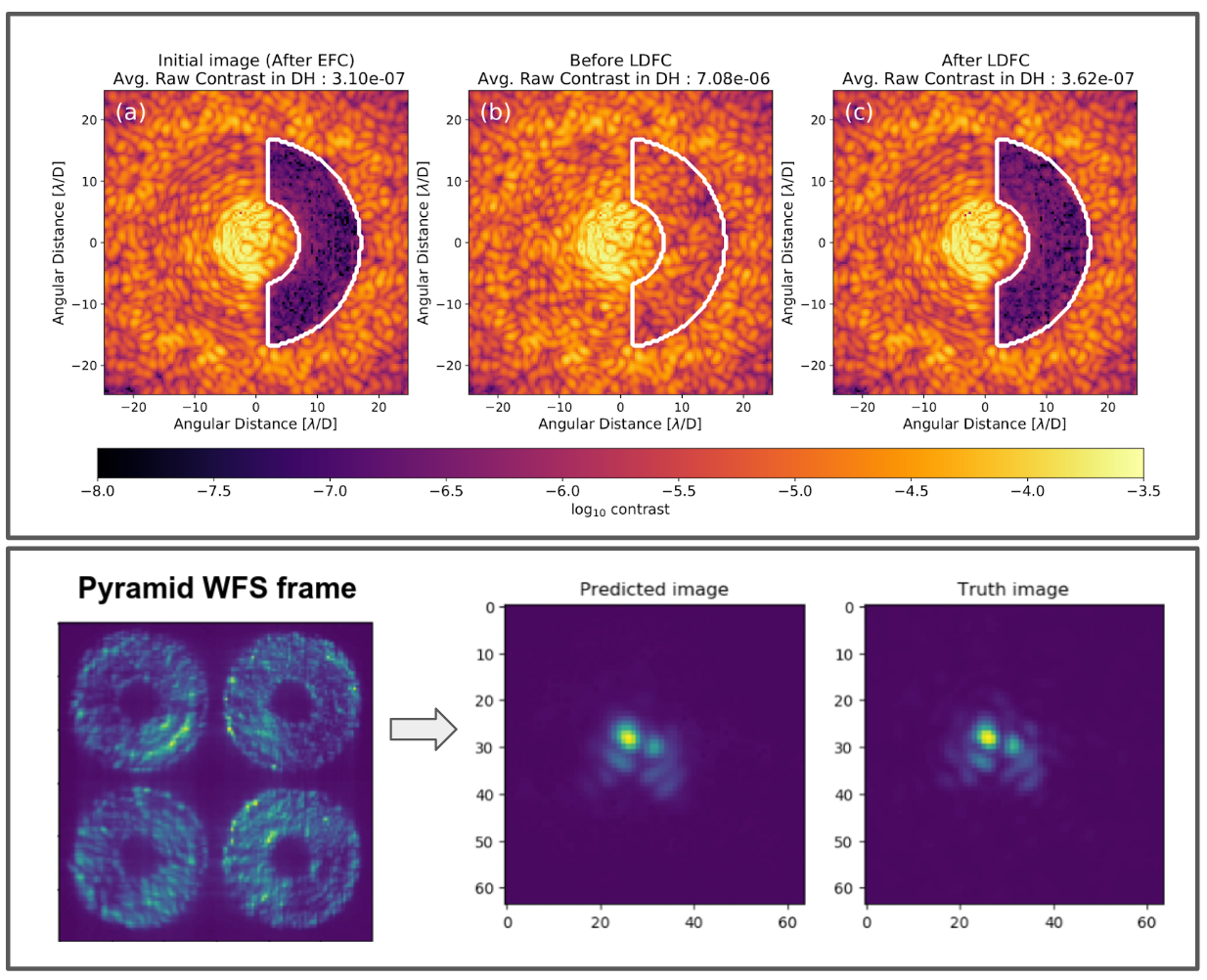}
    \vspace*{0.3cm}
    \caption{Wavefront control development: examples. Top: High-contrast imaging with focal plane wavefront sensing/control demonstrated on SCExAO using its internal source. Right: On-sky demonstration of PSF reconstruction using pyramid WFS telemetry.}
    \label{fig:HCIdev}
\end{figure}

The SCExAO system is supporting development of promising WFS/C techniques for high contrast imaging, including focal plane speckle control\cite{Ahn2022SPIE} and PSF reconstruction\cite{Guyon2022} as illustrated in Figure \ref{fig:HCIdev}. High frame rate science cameras in the system are also used for wavefront sensing \cite{Vievard2022LAPD, Deo2022}. Thanks to the AO188 upgrade to AO3k, these advanced modes of operation will be deployed on-sky to provide improved high contrast detection capabilities. SCExAO is also validating astrophotonics technologies for high angular resolution and high contrast imaging \cite{2021SPIE11823E..0CV}. This includes the GLINT and FIRST instruments that use coherent waveguides and photonic chips for high precision spectro-interferometric imaging, and new wavefront sensing concepts using dispersed interferometry \cite{Vievard2022, Norris2022GLINT} and Photonic Lantern \cite{Lin2022coupling,Norris2022PL}.

\acknowledgments 
 
This work is based on data collected at Subaru Telescope, which is operated by the National Astronomical Observatory of Japan. The authors wish to recognize and acknowledge the very significant cultural role and reverence that the summit of Maunakea has always had within the Hawaiian community. We are most fortunate to have the opportunity to conduct observations from this mountain. The authors also wish to acknowledge the critical importance of the current and recent Subaru Observatory daycrew, technicians, telescope operators, computer support, and office staff employees.  Their expertise, ingenuity, and dedication is indispensable to the continued successful operation of these observatories. The development of SCExAO was supported by the Japan Society for the Promotion of Science (Grant-in-Aid for Research \#23340051, \#26220704, \#23103002, \#19H00703 \& \#19H00695), the Astrobiology Center of the National Institutes of Natural Sciences, Japan, the Mt Cuba Foundation and the director's contingency fund at Subaru Telescope. KA acknowledges support from the Heising-Simons foundation.

\bibliography{report} 
\bibliographystyle{spiebib} 

\end{document}